# An Evaluated Certification Services System for the German National Root CA — Legally Binding and Trustworthy Transactions in E–Business and E–Government

A. Wiesmaier, M. Lippert, V. Karatsiolis, G. Raptis, J. Buchmann
Technische Universität Darmstadt
Department of Computer Science
64289 Darmstadt, Germany

*Abstract*— National Root CAs enable legally binding E–Business and E–Government transactions. This is a report about the development, the evaluation and the certification of the new certification services system for the German National Root CA. We illustrate why a new certification services system was necessary, and which requirements to the new system existed. Then we derive the tasks to be done from the mentioned requirements. After that we introduce the initial situation at the beginning of the project. We report about the very process and talk about some unfamiliar situations, special approaches and remarkable experiences. Finally we present the ready IT system and its impact to E–Business and E–Government.

## I. Introduction

The directive 1999/93/EC [1] of the European Union obligates each member to operate a National Root CA (NRC). Each one spans a national public key infrastructure (PKI). This enables legally binding digital signatures within each country. Such signatures are considered equivalent to regular hand-written signatures in most cases. They are called "qualified electronic signatures". This provides reliable and trustworthy E–Business and E–Government transactions. By having each NRC mutually "cross–recognizing" all others, this infrastructure is lifted to an international context.

According to §3 of the German digital signature act (SigG) [2] the German Regulatory Authority for Telecommunications and Posts[1] (RegTP) is responsible for operating the NRC of Germany. The RegTP launched a tender procedure to obtain a new IT system for its certification services in the first quarter of 2003. This was necessary because the old system could not easily be adapted to changes in the requirements. We now describe these requirements.

The reliability of digital signatures is based on cryptographic algorithms as public key signature schemes and cryptographic hash functions. As the research in cryptanalysis goes on the security status of the available cryptographic primitives degrades over the time. As an example Wang et al. have shown in [3] how to produce collisions for SHA1 better than with brute force. Lenstra et al. have shown in [4] how to produce colliding X.509 certificates. To reflect the developments the German Federal IT Security Agency[2](BSI) regularly suggests suitable cryptographic algorithms and associated parameters. The IT system must be easily adaptable to these guidelines.

In parallel, the ISIS–MTT [5] SigG–Profile [6] was advanced to version 1.1. It addresses technical requirements of the SigG and the respective ordinance (SigV) [7]. It is based on the X.509 [8] and the corresponding PKIX [9] specifications. Conformance to these standards was another demand.

In the old system, certificate revocation information was handled in a proprietary way based on black and white lists. They came along with a special viewer software and were not integrated into any common application. A demand of the tender procedure was to move to online certificate status checking based on the Online Certificate Status Protocol (OCSP) [10] and the LDAP Lightweight Directory Access Protocol [11].

Apart from these new requirements, the system had to comply to the SigG and the SigV. This included the certification of the overall security concept as well as the evaluation of parts of the system according to Com-mon Criteria (CC) [12] or the Information Technology Security Evaluation Criteria (ITSEC) [13], respectively. Furthermore, it must be possible to migrate the data assembled in the old system to the new one.

---

[1] http://www.regtp.de/en/index.html

[2] http://www.bsi.de/english/index.htm

The FlexSecure GmbH,[3] a spin-off from the Technische Universität Darmstadt,[4] was as a subcontractor of T–Systems[5] responsible for the software components of the NRC. This paper reports on this project from the developer's point of view.

## II. OUR TASKS

We had to participate in the development of the overall system design as well as the preparation of the respective security concept. The most challenging task was to integrate our software (FlexiTRUST) into the rest of the system in a way that suffices the requirements of SigG, SigV and ISIS–MTT and preserves as much of the flexibility of our software as possible. Therefore, we had to go into the very details of the respective documents to be able to consider their impact on our design.

We had to prepare the software for a successful accreditation. We therefore evaluated it to the required security level of the Common Criteria. This is described in Section IV and had several aspects. The software had to be adapted to higher security requirements. Evaluation of a system causes a lot of additional documentation to be written and most of it is not suitable for any other purposes than evaluation. Lastly, it meant diligent testing of the whole system. This was even more an issue, since we would not be able to change any lines of code afterwards without voiding the evaluation result. The testing had to be performed and documented in a formalized way.

We had to implement an OCSP responder that conforms to the respective requirements, most notably the ISIS–MTT SigG–Profile. This component had to be built from scratch and integrated into the existing system. This was even more an issue, since the component has to be highly available while producing OCSP responses with qualified electronic signatures (SigG) using evaluated signature creation devices (smart cards).

Formalizing the implemented security mechanisms was another requirement. According to the CC–methodology we had to mathematically prove that all cryptographic parameters (like passwords) are resistant against attacks with the formal potential "high". The logging had to be extended. The CC dictates which events have to be logged in the various functional components. We had to identify those in our software and implement and prove their audit proof logging. Also the delivery procedures and the installation procedures of the software had to be specified, described and documented in a formal way.

Lastly, the evaluated system had to be put into operation at the RegTP's site. This included setting up the software, performing integration tests, migrating the old data to the new system and training the RegTP's personnel.

## III. INITIAL SITUATION

When we started the project we already had a Trust Center (TC) software called FlexiTRUST. It is a component based system which supports all necessary tasks. These are registration, certification, certificate revocation, key generation and publication of the respective products. FlexiTRUST has been developed as a part of our academic work. Thus it had some experimental characteristics. The main design goals had so far been the flexibility to seamlessly exchange cryptographic algorithms and parameters, to easily integrate into existing workflows and to scale to ongoing load demands.

FlexiTRUST was already successfully employed in a variety of projects both in academic and business environments. Each project had different characteristics and requirements, but none of them were comparable to the project described here. This project had a lot of entirely new aspects which were not covered by the original design.

FlexiTRUST did not pass a certification process as demanded for an accredited certification services provider (CSP) before. It also had not been evaluated according to the CC, yet. It was evident, that most of the security functions of FlexiTRUST would have to be enhanced to achieve the high demands.

FlexiTRUST already complied to the X.509 and PKIX standards families. It supported all necessary and commonly used extensions of certificates and certificate revocation lists. It already provided mechanisms to integrate further extensions as demanded. Thus, being conformant to the ISIS-MTT SigG profile seemed to be straightforward.

The project schedule was extremely tight. The system had to be designed, adapted, tested, documented, evaluated, installed and certified within 6 months. As all parties (the developers, the evaluators, the certifiers and the customer) announced a high level of motivation and willingness for cooperation we were confident to succeed.

---

[3] http://www.flexsecure.de/ojava/home_en.html
[4] http://www.tu-darmstadt.de/index.en.html
[5] www.t-systems.com

## IV. Process

The most difficult task was to make FlexiTRUST fulfill the security requirements of the SigG and SigV within the given time. This became even more complex as we decided to evaluate the whole software instead of only the obligatory parts. This was done to be able to benefit from the evaluation in other projects. The approach was to take the security concept demanded by the SigG as the basis for deriving a security target (ST) for the evaluation process according to the CC. The FlexiTRUST system then was adapted accordingly and the necessary documentation was produced. The certification of the security concept of FlexiTRUST amounts to certifying the evaluation results. It soon became clear, that this approach would not succeed, if we took these steps sequentially as it is common practice.

The idea was to solve these tasks in parallel and in close cooperation with all parties. While the security target was developed together with the evaluating and the certifying authority, we already began to adapt our system to those requirements which seemed to be stable in the ST. Parallel to changing the software we started developing the high and low level design documents. Whenever we finished a document, we immediately consulted the evaluator. Thereby we obtained three major advantages. Firstly, the evaluating authority very early gained an outline as well as detailed information about the structure of our system. Secondly, we could take their responses into account for the development of the other documents. Thirdly, by linking the requirements of the security target very early to our implementation, we got a better understanding of them. The last two points were especially valuable, since we have not had any experience in evaluation processes.

The parallel work in close cooperation with the other parties was also valuable in the documentation process. It saved us from reading and understanding the entire CC framework and allowed us to concentrate on what was really important for our project. The first developer who wrote a certain kind of document had to learn this from the standards. The guidance from the other parties, as well as taking the common evaluation methodology (CEM) [14] into account, helped to avoid mistakes, misinterpretations and dead ends. This developer was then able to support the next team member with the same task. By this, knowledge spread quickly in the whole team. This speeded up the evaluation process considerably.

A crucial point was to decide, which security requirements to achieve solely by FlexiTRUST and which to delegate to the environment. A trade-of needed to be found between the time necessary for adaptations and the flexibility gained or lost by delegating them to the environment. As shown later in this section it was not always possible to implement the best solution due to the tight schedule. But particularly because of the close collaboration with the evaluating and certifying parties, we came to a working solution that can be considered optimal within the given context.

Being conformant to the ISIS–MTT specification turned out to be more complex than we thought at first. The upcoming version 1.1 was not yet stable. Furthermore, we had no client applications available to test against except for the test bed implemented by Secorvo Security Consulting GmbH.[6]

According to the requirements of the Common Criteria we had to defend our system also against attacks from inside the trust center that are carried out with the attack potential "high". This had by far the most striking impact on the adaptation process. It could not by any means be achieved or supported by the environment like e.g. the operating system. While the registration front-end already employed strong authentication and role based access control, this was not true for starting and stopping the trust center components as well as the protection of the secrets used within the processes (like e.g. database passwords). To counter the high attack potential we implemented strong authentication based on smart cards for all accesses to the software. To defend against attacks from inside we employed a multi party control scheme. In order to satisfy the CC requirement of establishing "trusted channels" we had to assure that even the end–points of the internal communication are authentic. This caused among other things, that we wrapped the entire communication between the publishing component and the LDAP servers in an SSL protocol and protected the required keys with the dual control scheme mentioned above.

The development process did not resemble what we had faced so far. Usually we had been working at our office or at home — most of the time being connected to the internet. According to CC the sources of the new FlexiTRUST version had to be highly protected against unauthorized access. Now, we had to work in a dedicated access controlled area and on dedicated machines. The source code repository was off–line and each change to it was monitored and had to be signed by the developer and the officer responsible for the respective module.

We planned to use an evaluated and accredited highly secure signature generation component from a third–party vendor. This component would include the card reader as well as the driver software. This would have allowed us to seamlessly exchange this component e.g. to adapt to different cryptographic algorithms. It turned out, that the driver software would not be available in an evaluated version in time. Thus we had to evaluate the driver as a component of our system. Apart from the extra work to have it evaluated, we also lost the ability to change this component without the reevaluation of our system.

---
[6] http://www.secorvo.de/

## V. RESULT

Figure 1 shows the main components of FlexiTRUST. The Registration Authority (RA) is responsible for collecting and verifying the end entities data. The Certificate Management Authority (CMA) provides the services dealing with the TC products. The Key Authority (KA) is sheltered by the RA and the CMA. Among other duties it signs the TC products with the issuer private keys. See [15] for more information on the KA.

Fig. 1
FLEXITRUST ARCHITECTURE

FlexiTRUST employs a generic provider architecture for dealing with cryptographic algorithms that abstracts from implementation specific details. This allows to seamlessly exchange the algorithms by installing a different cryptographic provider. If cryptographic hardware is involved the exchange of algorithms clearly is limited to those algorithms which are supported by the currently available cryptographic hardware. As we use the standard PCKS#11 [16] interface, such hardware can also be replaced. By default we deliver our own provider called FlexiPROVIDER[7] which provides a variety of established and novel algorithms.

[7] http://www.flexiprovider.de/

FlexiTRUST is multi client capable and therefore supports virtual hosting. This means that it is possible to host different, independent trust centers in parallel and at one site. The German NRC is therefore able to offer this as a service to the NRCs of other countries. Moreover, this allows to take over the duties of those certification service providers which ceased operation. It is the duty of NRCs to carry on the directory services of its deceased accredited certification service providers.

FlexiTRUST is built of components. Each component again consists of modules for the respective tasks that it has to perform. This allows integrating it into different environments by applying, configuring or omitting certain modules. It is further possible to distribute the installation over a network, either on the component level or even on the module level. This enables to scale the installation with respect to its load. If the load of the overall system is small, everything can be installed at one spot (e.g. on one laptop). As load increases the modules that are concerned can be replicated and/or offloaded to different machines in order to distribute the load. This was e.g. done for the OCSP responder. To be able to apply a qualified signature to the OCSP responses, the responder utilizes certified cryptographic hardware. In our case these are standard smart card readers with evaluated cards. A reader, a card and the required control logic (software) are implemented as one signer module. To scale the number of responses that can be signed in a certain amount of time, the number of active signer modules is simply increased.

There are two possibilities to run the system in a fault tolerant mode. One is to generate redundancy on module level as it is done for high load situations. If one module fails the others can take over its work. The second is to replicate components or even the entire trust center at a different spot. Depending on the components and on security considerations, they can be operated as hot or cold standby systems. In this project we decided to replicate the entire trust center. All modules that are concerned with the revocation of certificates (including OCSP and LDAP) are operated as hot standby systems, whereas all other modules run as cold standby systems. The two trust center instances are operated at two different cities. The hot standby systems are synchronized over a dedicated and protected line, the others via portable media (which increases the security).

The modularized approach also enables to adjust the security level. High security modules can be replaced by less secure ones. For example the OCSP's smart card signer module can be replaced by an implementation that uses keys held in software. There is a variety of reasonable combinations to reach different security levels.

Clearly exchanging components or modules forces a re–evaluation of the new configuration (if desired).

The generated products and offered services are fully standards compliant. The Certificates and Certificate Revocation Lists are following the PKIX standard respective the ISIS–MTT SigG profile. The offered services use standard protocols as LDAP or OCSP.

## VI. IMPACT

The introduced certification service system enables legally binding and globally trustworthy transactions in E–Commerce and E–Government.

FlexiTRUST is installed as a NRC. It can be installed as NRC for other countries, too. Further it is possible to host foreign NRC within existing installations. Thus, FlexiTRUST enables legally binding digital signatures in an international context. This is a basic requirement for global trustworthy E–Transactions.

FlexiTRUST is able to be used with arbitrary cryptographic algorithms and parameters. Thus it can be adjusted to the ongoing cryptographic efforts. In addition this makes the TC suitable to be used with the Fail–Safe–Concepts proposed by Maseberg [17]. Due to these two points it is possible to ensure long term security. Thus long term non–repudiation for E–Transactions is possible.

Having a set of NRCs alone is not enough. The customers / citizens and the companies / civil services have to be equipped with keys and certificates. This efforts that a small meshed infrastructure of PKIs has to be established. Those PKIs have to conform to the respective laws to provide an environment suitable for legally binding signatures. For small companies it might sufficient to deal with a few dozens of certificates. But the TCs of big corporate groups may have to deal with thousands of certificates and might have to answer millions of status requests. And the services have to be guaranteed. FlexiTRUST can easily be installed at companies, civil services and other institutions as it can easily be integrated in the existing workflows. It is certified to fulfill the legal demands. In addition it is possible to install it in the whole range between low load and high load cases. The robustness of the system is ensured by failover mechanisms. As we see, FlexiTRUST enables the area–wide participation of institutions and people in E–Business and E–Government.

Having all people and institutions participating in PKIs still is not enough. It must be ensured that each entity is able to communicate with any other entity. FlexiTRUST produces standard compliant products and offers standard compliant services. This guarantees a high interoperability and ensures that using the infrastructure is easy and efforts only minimal training of the participants. Thereby it is possible to have comfortable E–Transactions between any entities.

By having legally binding long lasting signatures, an area–wide participation, a high interoperability and a comfortable usage, the trust in the infrastructure as well as its acceptance is raised. This addresses the most important issue on E–Business and E–Government. The people must be willing to execute E–Transactions.

## VII. CONCLUSION

By winning the respective tender procedure we had the task to deliver the TC for the new German NRC which is hosted by the RegTP. This was a remarkable duty as we had to fulfill many special demands laid on us by the BSI, the ISIS–MTT standards, the SigG respectively the SigV and particularly by the obligatory CC–Evaluation procedure. For this we had to execute different tasks. We had to prepare the system for a successful accreditation. This included a successful evaluation. Finally we had to install the system on the target platforms and to configure the environment properly.

The most crucial point in the project was the extremely tight schedule. We had 6 month for fulfilling all the above mentioned tasks. To solve this we worked in close cooperation with the evaluating party, the certifying party and the customer. By this we were able to parallelize tasks which usually are executed as serial steps. Another remarkable point was that we had to conform to the upcoming and not yet stable version 1.1 of ISIS–MTT.

The methodology as well as the requirements of the CC in form and content had direct implications to the code and all other aspects of the project, such as the design, testing, configuration management and delivery of the software. Especially the security mechanisms we had to implement had to be conformant with the formal requirements of the CC.

Finally we succeeded in our task. In November 2003 FlexiTRUST 3.0 Release 0347 was evaluated according to CC EAL3 augmented. The strength of the established security mechanisms is "high". The CC certification report is confidential and thus can not be referenced here. In December 2003 the software was attested to conform to SigG and SigV. The respective certificate can be found in [18].

FlexiTRUST fulfills all legal demands, ensures long term security, is scalable in capacity, robust and produces standards conforming products. Thus it enables legally binding and trustworthy transactions in E–Business and E–Government.